\documentclass[a4paper,11pt,dvipsnames]{article}
\usepackage{jinstpub} 
\usepackage{lineno}
\usepackage{here}
\usepackage{subfig}
\usepackage{subcaption}
\usepackage{multicol}

\title{IOTA Experiment for Proton Pulse Compression at Extreme Space-Charge}

\author[a]{B.~D.~Simons,\note{Corresponding author.}}
\author[b]{N.~Banerjee}
\author[b]{J.~Eldred,}
\author[a]{and V.~Shiltsev}
\affiliation[a]{Department of Physics, Northern Illinois University,\\ DeKalb, IL, 60115, U.S.A.}
\affiliation[b]{Fermi National Accelerator Laboratory, \\ Batavia, Illinois 60510 U.S.A.}

\emailAdd{bsimons1@niu.edu}
\emailAdd{jseldred@fnal.gov} 

\abstract{The longitudinal compression of high-intensity, space-charge–dominated proton bunches is a critical requirement for future proton-driven muon colliders. We propose a proton bunch compression experiment at the Integrable Optics Test Accelerator (IOTA) storage ring at Fermilab to investigate optimal radio-frequency (RF) cavity parameters and lattice configurations. IOTA is a compact, fixed-energy storage ring dedicated to beam physics R\&D and capable of circulating a 2.5 MeV proton beam under extreme space-charge conditions. Using the ImpactX code with its 3D space-charge solver, simulations indicate that the bunch length can be rapidly reduced by at least a factor of two without appreciable degradation of transverse beam quality—even in the strong space-charge regime. However, longitudinal defocusing due to the space-charge remains a significant challenge in short-pulsed intense proton bunches, and the optimization of compression under these conditions is discussed.}

\keywords{Muon colliders, space-charge effects, proton beams, emittance evolution, bunch compression}

\begin{document}
\maketitle
\flushbottom

\section{Introduction}
\label{sec:intro}

Short high-intensity proton bunches are indispensable for a wide range of future projects, including spallation neutron sources, neutrino superbeams and factories, muon colliders, high-energy proton colliders, and plasma wakefield accelerators' drivers \cite{spl2008compress, lebedev2010proton, MC2014frontend,   assmann2009generation}.

The proton bunch length is determined by the longitudinal emittance at the source and by the details of acceleration and manipulation through the chain of linear and circular accelerators—typically operating at low RF frequencies and low gradients—and is often too long for the desired applications. The bunch length generally scales with the square root of the emittance and the fourth root of the RF voltage and frequency $V_{\rm RF} f_{RF}$ \cite{Syphers}, making equilibrium bunch lengths long, since the emittance is usually constrained by space-charge effects and the RF voltage by limitations of cavity length and cost.

Non-equilibrium bunch-compression techniques rely on introducing an energy difference between the head and tail of the bunch, which is then passed through a dispersive optical system—either a linear beamline or a circular section—where a difference in path lengths bring the head and tail closer together.

An appealing scheme for a single-pass bunch compression after extraction assumes sending the beam through an RF cavity close to the zero crossing of the RF wave, thus introducing a position-dependent energy, followed by a region with nonzero momentum compaction factor, such as a chicane. While this scheme works well for short, low- to medium-energy electron bunches (see, e.g., Ref. \cite{rossbach2014fels}), it is not very practical for proton bunches, which are usually much longer (and therefore require low-frequency RF) and at medium to high energies, at which the required low-frequency RF voltage becomes substantial and scales with the proton beam momentum.

Proton storage rings, in principle, allow bunch compression using a moderate RF system. For example, since the equilibrium bunch length in a proton storage ring scales inversely with the fourth root of the RF voltage and with the fourth root of the slippage factor $\eta \equiv (\alpha_c - 1/\gamma^2)$ (where $\gamma$ is the relativistic Lorentz factor), adiabatically increasing the RF voltage or adiabatically reducing the momentum compaction $\alpha_c$ prior to extraction can lead to bunch compression.
A more sophisticated variant involves applying a series of successive up and down changes (typically in $V_{RF}$) at twice the synchrotron frequency over one or several synchrotron periods prior to extraction. This is a standard operation when transferring beams between synchrotrons (e.g., PETRA to HERA, PS to SPS, Fermilab Booster to Main Injector \cite{assmann2009generation, eldred2021beam}).

Unfortunately, such adiabatic or semi-adiabatic manipulations are not suitable for beams whose dynamics are strongly dominated by space-charge effects, as is the case in proton rings for a neutrino factory or a muon collider \cite{ankenbrandt1999status, alexahin2012design, autin2000analysis, ng2013space, thomason2013proton}. Strong space-charge forces can cause unacceptable emittance blow-up, bunch lengthening, and particle losses within just a few dozen turns. This necessitates correspondingly fast RF manipulations, sufficient RF voltage to overcome space-charge forces, and beam energies far enough from transition so that the final bunch rotation can occur rapidly.

For instance, in a {\it snap bunch rotation} within a storage ring, a rapid increase in the RF voltage $V_{RF}$ can mismatch the RF bucket, making the initial bunch length too long for the new bucket shape. The bunch then executes synchrotron oscillations around a new equilibrium, and after a quarter of a synchrotron period, it reaches a minimum length, at which point it can be extracted.

\begin{figure}[!htb]
    \centering
    \includegraphics*[width=\columnwidth]{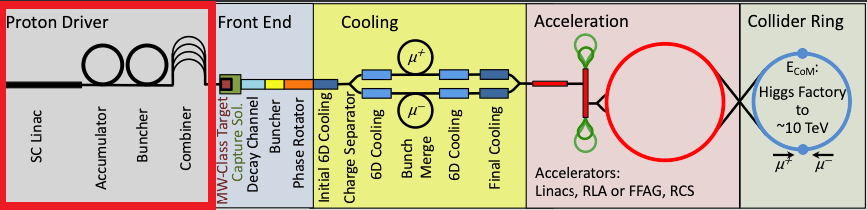}
    \caption{Schematic layout of the Muon Collider complex (adapted from~\cite{boscolo2019future}). 
}
    \label{fig:MuCLayout}
\end{figure}

A Muon Collider (MuC) is commonly considered a compelling option for a 10 TeV parton center-of-momentum (pCM) energy frontier collider \cite{P5, nas2025epp}, and Fermilab considered as a potential host facility for MuC R\&D projects as well as the collider itself \cite{black2024muon, stratakis2025essay}. The future MuC complex \cite{ankenbrandt1999status,  boscolo2019future, accettura2023towards} --  see Fig.~\ref{fig:MuCLayout} -- will be based on the 2–4 MW proton source:  H$^{-}$ ions are accelerated to 4–8 GeV in a superconducting linac, and then collected in the Accumulator Ring as a few intense proton bunches. The Buncher Ring (also called Compressor Ring) of the proton driver compresses each these into $\sim$3~ns long bunches which are sent to the Combiner so that all bunches converge on a pion production target. The pions are captured and allowed to decay into muons, which are then cooled, accelerated, and ultimately collided. 

Strong space-charge forces in short, high-intensity proton bunches can cause beam emittance growth, bunch lengthening, and particle losses (see, e.g.,~\cite{eldred2021beam}), thereby imposing a major performance constraint on the Muon Collider Accumulator and Combiner Rings~\cite{JeffIPAC,SophiaIPAC}. Accelerator R\&D aimed at producing short, intense proton pulses is not only essential for the realization of a future MuC, but also broadly relevant to next-generation high-energy physics (HEP) experiments, including dark sector searches, neutrino programs, and precision flavor physics~\cite{MuonWSR,BeamDump}.

The work presented here aims to develop a new R\&D program for bunch compression at Fermilab’s Integrable Optics Test Accelerator (IOTA)~\cite{Jinst} storage ring, to be carried out during its upcoming proton runs. FAST/IOTA is the only U.S. facility dedicated to intensity frontier accelerator R\&D \cite{abp2023roadmap}. As such, its research portfolio includes advanced techniques for mitigating coherent instabilities and space-charge effects, including nonlinear integrable optics \cite{DNlattice} and electron lenses \cite{shiltsev2015electron, stern2021self, elenses}. With current research program at IOTA during proton runs focused on achieving high beam intensity and brightness in proton rings, a natural synergy arises in extending this work toward compressing intense proton beams into short bunches — an essential capability for next-generation HEP experiments.

\begin{figure}[!htb]
    \centering
    \includegraphics[width=0.75\columnwidth]{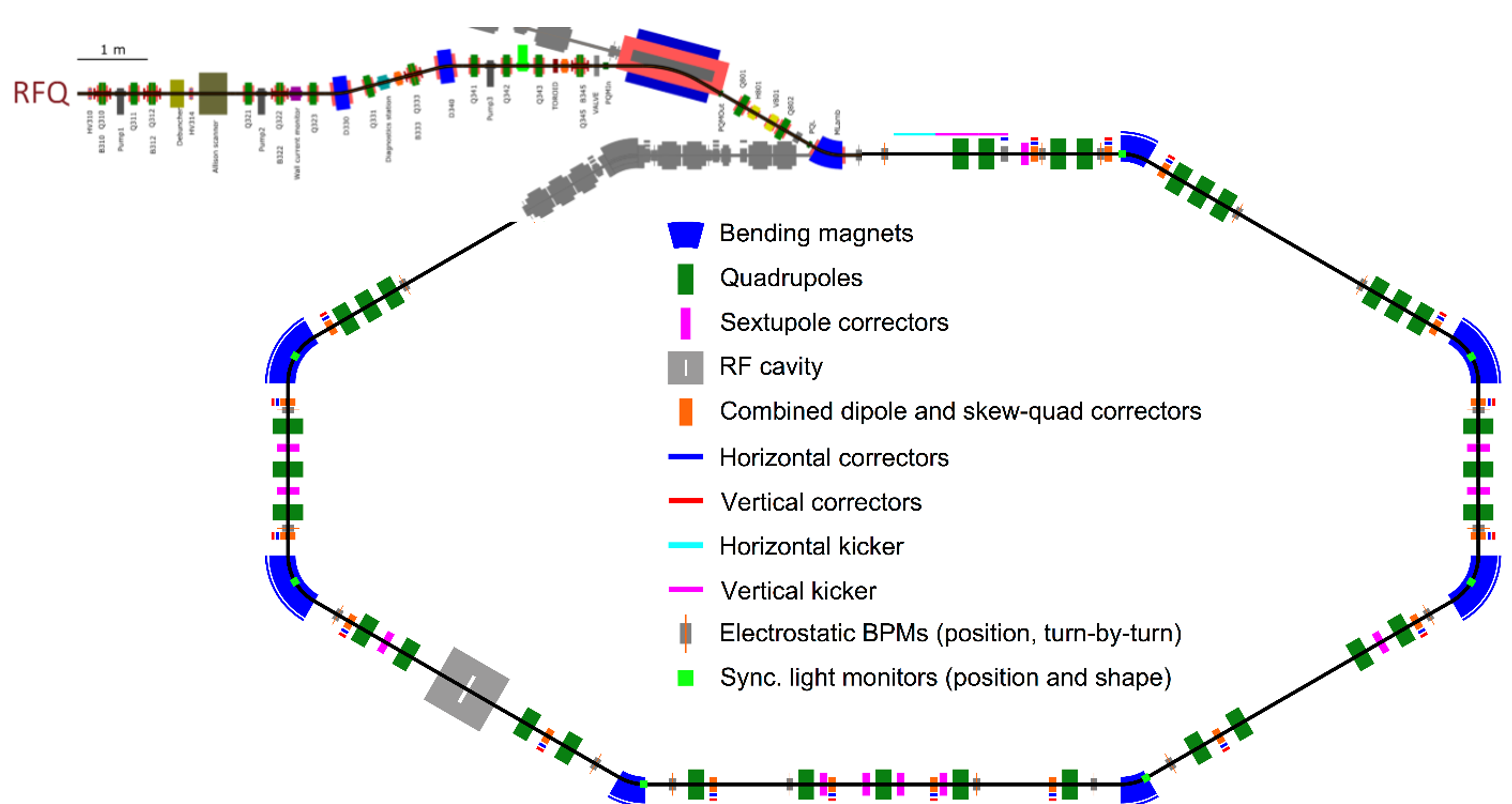}
    \caption{IOTA storage ring and proton injector line layout (from \cite{Jinst}).}
    \label{fig:IOTAring}
\end{figure}

The IOTA ring itself, shown in Fig.~\ref{fig:IOTAring} along with the proton injector line, is a compact, 40-m circumference fixed-energy storage ring that can circulate a 2.5 MeV proton beam with a beam current of up to 8 mA \cite{Jinst} and a corresponding incoherent space-charge tune shift $|\Delta Q_{sc}|$ of up to 0.5. The proposed accelerator R\&D program will incorporate factor of two bunch compression to extend the beam physics program to an even more extreme space-charge regime.

The IOTA 2.5 MeV proton injector (IPI) is currently being commissioned \cite{IPI} for the upcoming proton runs. 
As discussed in \cite{RF-Cap}, the RF cavity in the DR section of IOTA contains two RF gaps operating at harmonic numbers $h=4$ (2.19~MHz), which is used to fully bunch the proton beam, and $h=56$ (30.6~MHz), which is used to introduce longitudinal modulation for instrumentation. The design parameters for proton operations are shown in Table~\ref{tab:IOTAParams} and the various RF parameters for the different systems will be discussed in later sections.

The R\&D program, FAST/IOTA Bunch Rotation Experiment (FIBRE), constitutes an upgrade to the existing capabilities of the IOTA ring. Its primary goal is to achieve compression of intense proton beams into short pulses using the RF technique of \emph{snap bunch rotation}.

\begin{table}[!htb]
\centering
\begin{tabular}{|l | c | c |}\hline
   \textbf{Parameter} & \textbf{Value} & \textbf{Units}\\ \hline
    Kinetic energy $(\gamma - 1)mc^2$  & 2.5 & MeV\\ 
    Momentum $p$  & 68.5 & MeV/c \\ 
    Relativistic $\beta = v/c$ & 0.073 & \\ 
    Circumference $C$ & 39.97 & m \\
    Proton revolution time $T_0$  & 1.83 & \textmu s \\ %
    Phase-slip factor $\eta$ & -0.9256 & \\ \hline
        Revolution frequency $f_{0}$  & 0.5467 & MHz \\ 
    RF frequency $f_4$ at $h=4$  & 2.19 & MHz \\ %
    RF bucket length ($h$=4)  & 475 & ns \\
    RF voltage at $h=4$ $V_4$  & $0.65-10.3$ & kV \\ %
    Synchrotron tune $Q_s$ & $(0.87-3.5) \cdot 10^{-2}$ & \\
    Bare linear lattice tunes ($Q_x, Q_y$) & 5.3, 5.3 & \\ \hline
    Number of particles (max) $N_p$, $I_p$  & $9\cdot 10^{10}$  &  \\
        Max. beam current $I_p$  & 8  & mA \\
    Norm. RFQ beam emittance $\varepsilon_{N, x/y}$  & 0.3 & \textmu m \\
    R.m.s. relative momentum spread $dp/p$ & 0.001 & \\ 
    R.m.s. energy spread $\sigma_{E}$  & 5.0 & keV \\ \hline
\end{tabular}
\caption{Main parameters of the IOTA Ring operating with protons.}
\label{tab:IOTAParams}
\end{table}

\section{Proposed Experiment}
\label{sec:experiment}


The FIBRE experiment starts with a 2.5~MeV proton beam from a 325~MHz RFQ that is injected into IOTA, which will operate with little or no RF voltage. After a few turns, the initial bunch structure will decohere, and a coasting proton beam will fill the entire IOTA circumference. The voltage of the IOTA $h=4$ RF system will then be adiabatically ramped up to its maximum capture value. Shortly thereafter, the RF voltage will be rapidly increased to induce longitudinal phase space rotation of the proton bunch, reaching minimum bunch length at one-quarter of the synchrotron period (\emph{snap bunch rotation}). All relevant proton beam parameters—such as current, bunch length, transverse beam sizes, and emittances will be recorded using IOTA's beam diagnostics systems and analyzed to optimize the bunch compression ratio and benchmark the results against detailed computer simulations of the IOTA beam dynamics.


Questions we look to answer with this research include understanding how the transverse and longitudinal beam parameters evolve during these RF manipulations across a range of beam currents—from the lowest detectable to the highest achievable from the RFQ injector. We will determine the extent to which longitudinal and transverse space-charge forces govern the evolution of beam properties, including final bunch length, emittance growth, and particle losses. In turn, we will determine if effective optimization or mitigation techniques can be applied to control beam degradation and further enhance the compression of high-intensity proton bunches. 

\subsection{RF Manipulations}

In the original scheme for proton operation~\cite{Jinst}, the voltage of the 2.19~MHz RF cavity gap is limited to 1~kV as a consequence of the power available from the planned IOTA high-level RF system. This voltage is sufficient for bunching the proton beam, but not adequate for the RF manipulations required for achieving shorter bunch length. The benefit of a higher RF voltage would not only be a cornerstone of a bunch compression experiment, but enhance the longitudinal performance for the entire intense beam research program. For snap bunch rotation, the system would also need to provide sufficient power to step up the voltage over just a few \textmu s. An appropriate RF cavity design has been identified, and for the simulation analysis that follows we will assume such a broadly capable RF system will be used. 

The 325~MHz radio frequency quadrupole (RFQ) \cite{RFQ} in the IPI accelerates low energy protons to 2.5~MeV resulting in a train of bunches, each with rms length of 0.3~ns, spaced apart by 3.1~ns. This bunch train is injected into IOTA in a single turn and the beam fills the entire circumference of the ring. At this low energy and with the anticipated momentum spread, the off-momentum protons will slip by $\Delta T \approx T_0 |\eta| \Delta\delta = 2.8$~ns on completing a turn in the ring. Consequently, the beam debunches and effectively becomes a coasting beam after a few turns. Neglecting space charge, the minimum voltage required to capture the protons into a bunch is\cite{Syphers} 
\begin{equation}\label{eq1}
    V_{0} = \left(\frac{1}{e}\right)\frac{(n\sigma_{E})^2 \pi^3 h |\eta|}{8\beta^2 E_{s}}
\end{equation}
where $n$ is the number of standard deviations of energy acceptance ($n\sigma_E$) we want for the injected beam, $\eta$ is the phase slip-factor of the lattice, $\beta$ is the relativistic $\beta$ from Table~\ref{tab:IOTAParams}, and $E_s$ is the total energy of the synchronous particle. 
In order to avoid filamentation in longitudinal phase-space and undesirable emittance growth, the capture voltage will be adiabatically ramped to some maximum value. For modeling the adiabatic capture process, we used an optimized adiabatic capture curve between an initial voltage, $V_0$, and a final voltage, $V_1$, given by \cite{KYRF},
\begin{equation}\label{eq2}
    V(t)=\begin{cases}
			\frac{V_0}{\left[1-\left(1-\sqrt{\frac{V_0}{V_1}}\right) \frac{t}{t_1}\right]^2} & ;\text{$t < t_1$}\\
            V_1 & ;\text{$t \geq t_1$}
		 \end{cases}
\end{equation}
where
\begin{equation}\label{eq3}
    t_{1} = \left(1 - \sqrt{\frac{V_0}{V_1}}\right)\frac{n_{ad}}{2\pi \nu_{s,0}}
\end{equation}
with $t_1$ providing the time required for the capture ramp in terms of number of turns. For all analysis, the adiabatic capture begins with an initial cavity voltage $V_0=10$~V and the adiabaticity number, $n_{ad}$, was set to 10. Future work does involve further optimization of the capture process as a function of the injected current. 

\subsection{Snap Bunch-Rotation}
\begin{figure}[!htb]
    \centering
    \includegraphics*[width=0.75\columnwidth]{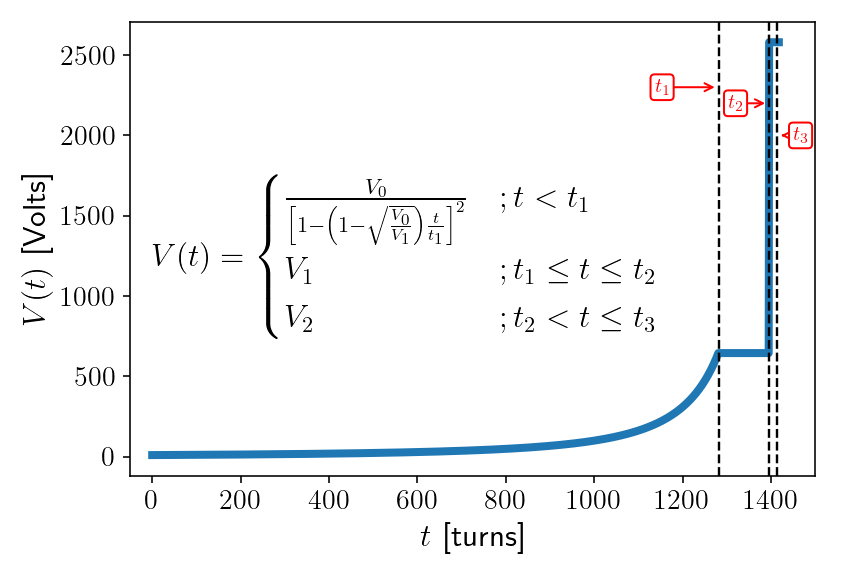}
    \caption{Time evolution of RF cavity voltage: during adiabatic capture ($0 - t_1$), stationary bucket to allow beam to reach equilibrium ($t_1 - t_2$), and finally snap rotation ($t_2 - t_3$).}
    \label{fig:RF}
\end{figure}

Snap bunch-rotation involves instantaneously increasing the RF voltage in the cavity, which expands the bucket height and shortens the synchrotron period, and then allowing the bunch to rotate in longitudinal phase space for 1/4 of a synchrotron period. In the application, the compressed bunch is extracted, while in the FAST/IOTA experiment the beam quality will be analyzed at the bunch length minimum. \\
\indent
The ratio between the initial (in our case after adiabatic capture) bunch length, and the compressed bunch length is called the compression factor, $r_c$. We find the required voltage for snap bunch-rotation of the core of the beam through the relation
\begin{equation}\label{eq4}
    r_c = \sqrt{\frac{V_{rot}}{V_{cap}}}
\end{equation}
For capturing and bunching $3\sigma$ of the beam's initial momentum spread, a voltage of $V_{cap} = 645$~V is required. With a desired compression factor of at least 2, we find that the required snap voltage is $V_{rot} = 2.58$~kV. The voltage provided by the RF cavity from the start of adiabatic capture to the end of snap rotation for a compression factor of two ($r_c=2$) is detailed in Fig.~\ref{fig:RF}. 
Neglecting any intensity driven effects, the process of adiabatically bunching/capturing the beam happens over $1282$ turns, or $\sim 2.34$~ms, the beam then takes $114$ turns (1 synchrotron period) to equilibrate in a stationary RF bucket at the same final capture voltage ($645$~V) before the snap rotation takes place, which requires just $17$ turns for full compression.

\subsection{Diagnostics in IOTA}
The initial diagnostics available include a DC current transformer (DCCT) which provides total beam current as well as some insight into overall beam stability by detecting losses that may occur. There is also a wall current monitor in the ring, which should provide the longitudinal profile of the beam. The beam centroid's transverse position will be measured with beam position monitors (BPMs), though the beam may need to be modulated using the 30.6~MHz (h=56) RF gap in order for the BPMs to detect it. 
Ionization Profile Monitors (IPMs) are under construction and will be installed in IOTA, enabling turn-by-turn measurements of the beam’s transverse profile ~\cite{IPM}. 

\section{Simulations with Space Charge}
The experiment is being simulated in ImpactX \cite{impactx}, which is an $s$-based beam dynamics code with space charge.  The space-charge solver is fully 3D and uses an integrated Greens function space charge solver ("FFT solver") that assumes open boundary conditions at the edges of the simulation domain. The work presented here uses a simulation domain that spans the maximum and minimum of the beam in each dimension with a 5\% margin. 

\subsection{Overview of Space Charge Effects}
As a reminder, space charge forces are a direct effect of a distribution's self-fields which depend on how the charge itself is distributed. 
The transverse incoherent space charge tune shift parameter, $|\Delta Q_{sc}|$, which is often a figure of merit for space charge studies, is given as  \cite{DQs}:
\begin{equation}\label{eq5}
        |\Delta Q_{sc}|_{x,y} = \frac{r_p N_p }{2\pi\beta_{rel}^2\gamma_{rel}^3} \frac{C}{h \sqrt{2\pi} \sigma_z} \left \langle {\beta_{x,y} \over \sigma_{x,y}(\sigma_x+\sigma_y)} \right \rangle
\end{equation}
where $N_p$ is the total number of protons in the ring, $r_p$ the classical radius of the proton, $C$ is the circumference of the ring, $\beta_{x,y}$ is the Courant-Snyder amplitude function, $\beta_{rel}$ and $\gamma_{rel}$ the relativistic Lorentz factors, $h$ is the RF harmonic, $\sigma_z$ is the rms bunch length, in meters, and $\sigma_{x,y}$ is the transverse rms beam size, in meters. The angled brackets in Eq.~\ref{eq5} implies taking the average over the circumference of the ring. Although Eq.~\ref{eq5} takes the absolute value of $\Delta Q_{sc}$, it is always negative, meaning the transverse tunes of the lattice will be depressed in the presence of significant space charge. The transverse rms beam sizes, $\sigma_{x,y}$, including any dispersive effects, are found as:
\begin{equation}\label{eq6}
  \sigma_{x,y} = \sqrt{\epsilon_{x,y} \beta_{x,y} + D_{x,y}^2\sigma_{\delta}^2}
\end{equation}
Where $\sigma_{\delta}=\Delta p/p_0$ is the relative momentum deviation, and $D_{x,y}$ are the horizontal and vertical dispersions.

From Eq.~\ref{eq5} it follows that longitudinal compression of the bunch increases the transverse incoherent space‑charge tune shift.  Consequently, particles in the core of the bunch are driven closer to, or even across, potentially dangerous resonances—most notably the integer resonance. As the core particles experience diffusion, the transverse beam size may expand to the point where particles are intercepted by the machine aperture, leading to beam loss.\\
\indent While transverse space-charge is always defocusing, the space charge effects seen longitudinally can be either defocusing or focusing, depending on the slip-factor of the beam, $\eta$. For IOTA's proton operation with negative $\eta$, the longitudinal space charge force is defocusing and the voltage ($V_{SC}$) from this defocusing force can be found as
\begin{equation}\label{eq8}
    V_{SC} = E_{s,SC}C_0
\end{equation}
where $E_{s,SC}$ is the electric field created by the space charge potential, acting in the direction of motion of the bunches, and $C_0$ is the circumference of the ring. $E_{s,SC}$ is given as \cite{NgUSPAS} 
\begin{equation}\label{eq9}
    E_{s,SC} = -\frac{e}{2\pi \beta c}\left|\frac{Z_0}{n}\right|_{SC} \frac{\partial \rho (\tau)}{\partial \tau}
\end{equation}
with $\left|\frac{Z_0}{n}\right|_{SC}$ being the longitudinal space charge impedance and for simplicity, we are assuming the longitudinal particle density to be a Gaussian distribution, 
\begin{equation*}
\rho (\tau) = \frac{N_b}{(\sqrt{2\pi}\sigma_\tau)}e^{-\frac{\tau^2}{2\sigma^2_\tau}}
\end{equation*}
with $N_b$ as the number of protons per bunch, $\sigma_\tau$ the rms bunch length in nanoseconds, and $\tau$ the length of the bucket in nanoseconds. Plugging everything back into Eq.~\ref{eq8}, we can approximate the longitudinal space charge potential as 
\begin{equation}\label{eq9}
    V_{SC} = \left(\frac{eN_b\tau Z_0g_0}{4\pi \sqrt{2\pi}\ \omega_0\gamma_{rel}^2 \beta_{rel}\sigma_\tau^3}\right)
\end{equation}
with $Z_0=376.73$~Ohms being the vacuum impedance, the geometry factor $g_0=1$, and $\omega_0  = 2\pi f_0$. A numerical example of Eq.~\ref{eq9} for a $4$~mA beam with a factor of two bunch compression illustrates the magnitude of the defocusing effect. At $4$~mA, $N_b=1.1418\times10^{10}$, we have a maximum space charge potential of $V_{SC}=131$~V by the end of adiabatic capture, compared to the maximum voltage of $645$~V provided by the RF cavity during this stage. By the end of snap rotation, that potential rises to $V_{SC}=1050$~V, compared to the $2580$~V coming from the cavity. The defocusing effects of the longitudinal space charge forces lead to a lengthening of the bunch and a smearing of the core of the beam. 

\subsection{Simulation Setup}

\begin{figure}[!htb]
    \centering
    \includegraphics[width=\columnwidth]{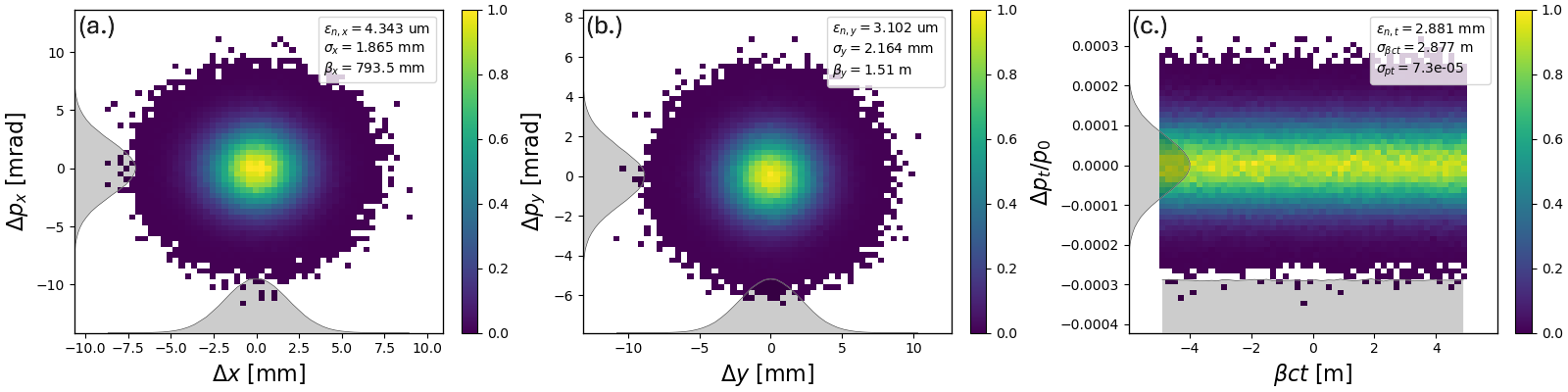}
    \caption{2D projections of the initial 6D distribution of the beam along with 1D projections on each axis. Panel (a.) is the horizontal phase space $\Delta p_x - \Delta x$, panel (b.) is the vertical phase space $\Delta p_y - \Delta y$, and panel (c.) is the longitudinal phase space $\Delta p_t$\protect\footnotemark$/p_0 - \Delta \beta ct$. The color indicates density, which has been normalized to 1.}
    \label{fig:InitDist}
\end{figure}

After some convergence studies of the overall setup parameters within ImpactX, the simulations proceeded with a [$512\times512\times128$] grid (x,y,z), $5\times10^6$ macroparticles, and 333 space charge kicks for one turn around the ring. The grid size is notably very fine in the transverse dimensions and this is to ensure an accurate representation of what is happening in the core of the beam. 
Shown in Fig.~\ref{fig:InitDist} is the initial 6D phase-space distribution used for all simulations. The initial distribution is assumed to be a Gaussian distribution in all transverse dimensions $(x, p_x, y, p_y)$, matched to the entrance of the bare lattice. The longitudinal phase-space consists of uncorrelated distributions - Gaussian in $p_z$ and uniform in $z$ - emulating an initially unbunched proton beam occupying the entire IOTA ring. 

The process of adiabatic capture is simulated and then the beam is allowed to equilibrate for one synchrotron period before the snap rotation takes place, as illustrated in Fig.~\ref{fig:RF}. All of the computationally intensive ImpactX simulations were run on a compute cluster called METIS, which is housed at Northern Illinois University's Center for Research Computing and Data \cite{METIS}.

Simulations were run for injected currents of 1~\textmu A, 0.5~mA, 1~mA, 2~mA, 4~mA, and 8~mA. The lowest current (1~\textmu A) serves as the reference case with very little space charge, the highest current (8~mA) is the quoted maximum current the injector can provide, and everything in between is meant to cover what the injector will actually output once commissioning is complete. For each injected current, simulations were ran for two different factors of compression; $r_c=2$ requiring a snap rotation voltage of $2.58$~kV and $r_c=4$, which requires a snap voltage of $10.32$~kV. Most of the results that follow showcase the $r_c=2$ scenario with $r_c=4$ comparisons included towards the end of the analysis. 

Particle losses were tracked at the most basic level, meaning that only apertures representing the vacuum beam pipe (two inch diameter) were included in all simulations. Realistic apertures for each element in the lattice will be included in future works. 

\subsection{Results \& Analysis}
\begin{figure}[!htb]
    \centering
    \includegraphics[width=0.75\columnwidth]{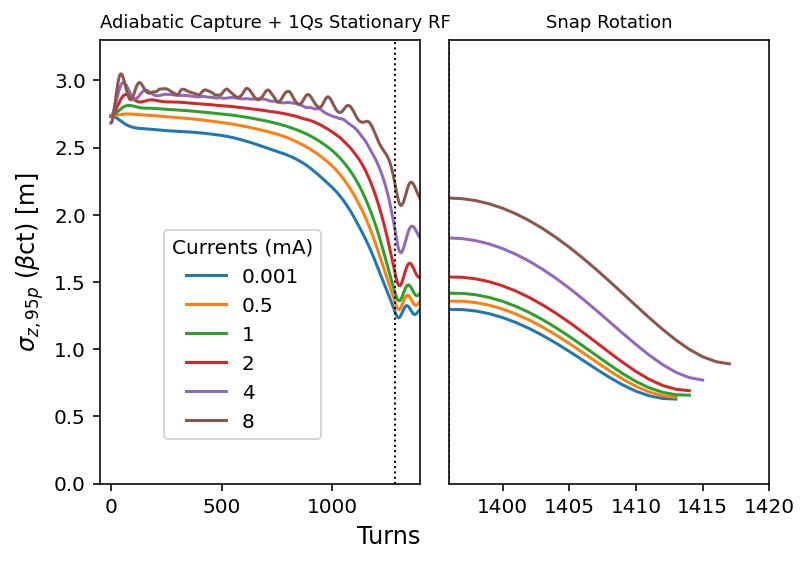}
    \caption{Evolution of the longitudinal rms bunch length $\sigma_{z,95\%}$, computed over adiabatic capture, a period of stationary RF, and then snap rotation. The vertical dotted line in the section on the left marks the transition from adiabatic capture/bunching to stationary RF, followed by snap rotation, which occupies the section on the right.}
    \label{fig:BL_all}
\end{figure}

\footnotetext{Here, $\Delta p_t$ is the deviation from the reference energy.}
The plot in Fig.~\ref{fig:BL_all} tracks $\sigma_{z,95\%}$, in meters, as a function of turn number for the range of injected currents mentioned previously, starting with 1~\textmu A and ending with 8~mA. Throughout this analysis, $\sigma_{z,95\%}$ refers to the rms bunch length of only the central 95\% of the distribution, providing a rough representation of the core of the bunch. 
\\
During capture, higher currents exhibit greater bunch lengthening due to longitudinal space-charge defocusing. In the final snap rotation phase, the number of turns required to reach minimum bunch length increases with current, reflecting the intensity-dependent elongation of the synchrotron period. Notably, the 8~mA curve shows both the largest initial length and the slowest compression, underscoring the impact of longitudinal space charge on the beam dynamics. During bunch rotation, the achieved bunch compression of $\sigma_{z,95\%}$ is actually slightly stronger than a factor of two, which is an effect of nonlinearities in the RF bucket on moderate tails while excluding the 5\% most extreme particles.

\begin{figure}[!ht]
    \centering
    \subfloat{{\includegraphics[width=0.49\columnwidth]{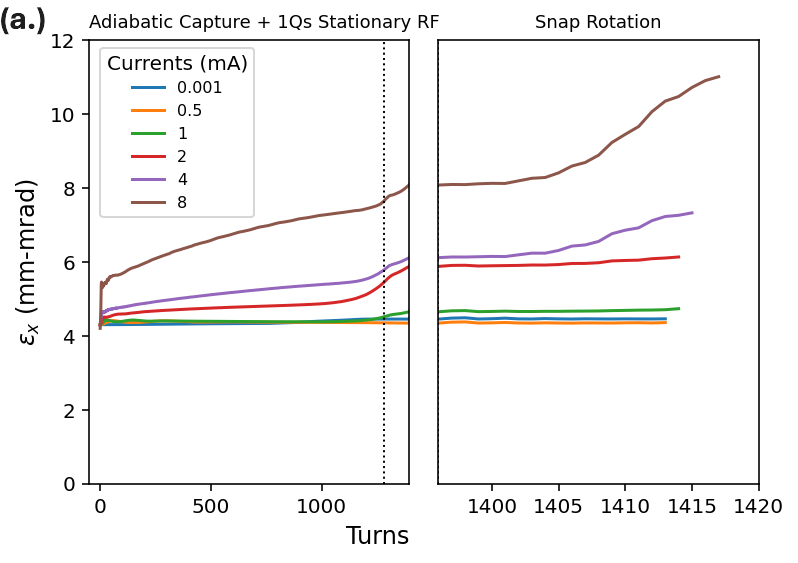} }}%
    \subfloat{{\includegraphics[width=0.49\columnwidth]{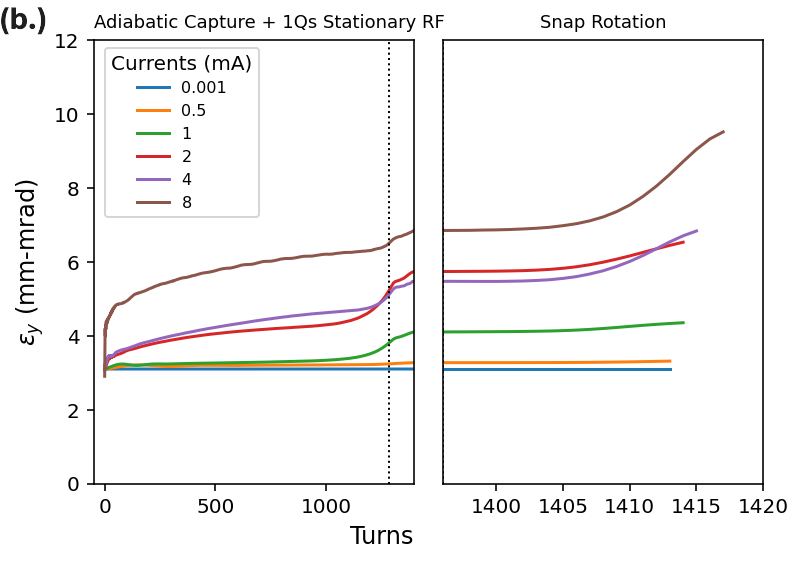} }}%
    \caption{Transverse rms emittance evolution in the horizontal (a.) and vertical (b.) planes, corrected for dispersion, for the same set of simulations shown in Fig.~\ref{fig:BL_all}.}%
    \label{fig:Emits}%
\end{figure}

Figure~\ref{fig:Emits} shows the evolution of the transverse beam emittances for the same simulation cases shown in Fig.~\ref{fig:BL_all}. In all cases, the initial transverse optics are kept the same, where any space-charge induced mismatch of the injected beam rapidly decoheres and the ultimate transverse beam emittance is driven by the capture and bunch rotation process.
The longitudinal bunch length evolution shown in Fig.~\ref{fig:BL_all} directly influences the transverse emittance behavior observed in Fig.~\ref{fig:Emits}. Along with the rapid decoherence from injected beam mismatch, as the beam current increases, the longitudinal space-charge defocusing force stretches the bunch during adiabatic capture, contributing to a larger initial bunch length. This elongation reduces the peak line charge density temporarily, but as the RF voltage ramps and the bunch compresses, the density of the core of the bunch rises in turn. The resulting increase in transverse space charge forces drives emittance growth, particularly during the snap rotation when the bunch length contracts rapidly. The steep transverse emittance growth coincides with the point when the highest charge density is achieved longitudinally. 

\begin{figure}[!h]
    \centering
    \subfloat{{\includegraphics[width=0.49\columnwidth]{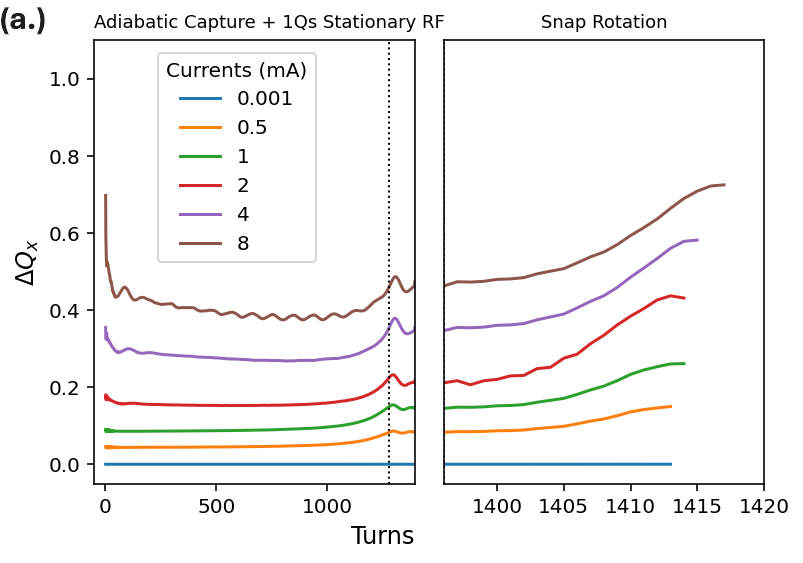} }}%
    \subfloat{{\includegraphics[width=0.49\columnwidth]{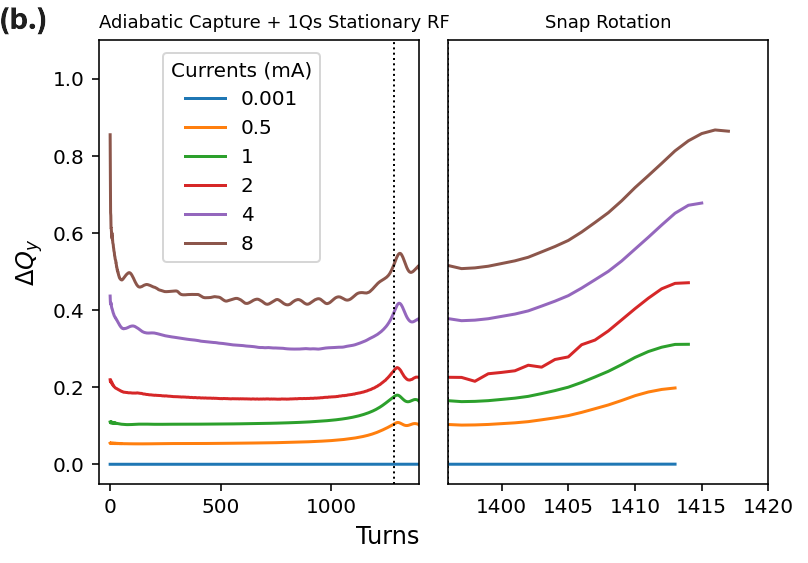} }}%
    \caption{Evolution of the incoherent SC tune shift parameter from Eq.~\ref{eq5}, for the range of proton beam currents, calculated using the bunch lengths and transverse sizes provided by the ImpactX simulations.}%
    \label{fig:DelQSC}%
\end{figure}

This interplay between longitudinal compression and transverse emittance growth is further reflected in the evolution of the incoherent space-charge tune shift shown in Fig.~\ref{fig:DelQSC}. The tune shift parameter being tracked in Fig.~\ref{fig:DelQSC} is calculated using Eq.~\ref{eq5} with the bunch lengths and transverse beam sizes from the ImpactX simulations shown in Fig.~\ref{fig:BL_all} and Fig.~\ref{fig:Emits}.

Figure~\ref{fig:DelQSC} reveals how transverse tune shifts evolve in response to longitudinal compression and the consequent increase in space charge forces. During capture, both $\Delta Q_x$ and $\Delta Q_y$ grow steadily, with higher currents exhibiting stronger tune depression. Once snap rotation begins, the tune shifts rapidly increase as the bunch contracts. The final space charge tune shift shown in Fig.~\ref{fig:DelQSC} is immense, however, because the snap rotation occurs rapidly (less than 22 turns or $\sim$ 30~\textmu s), the beam does not have time to reach a new equilibrium. Consequently the achieved tuneshifts are far greater than what is achieved without snap rotation. 

\begin{figure}[!t]
    \centering
    \subfloat{{\includegraphics[width=0.49\columnwidth]{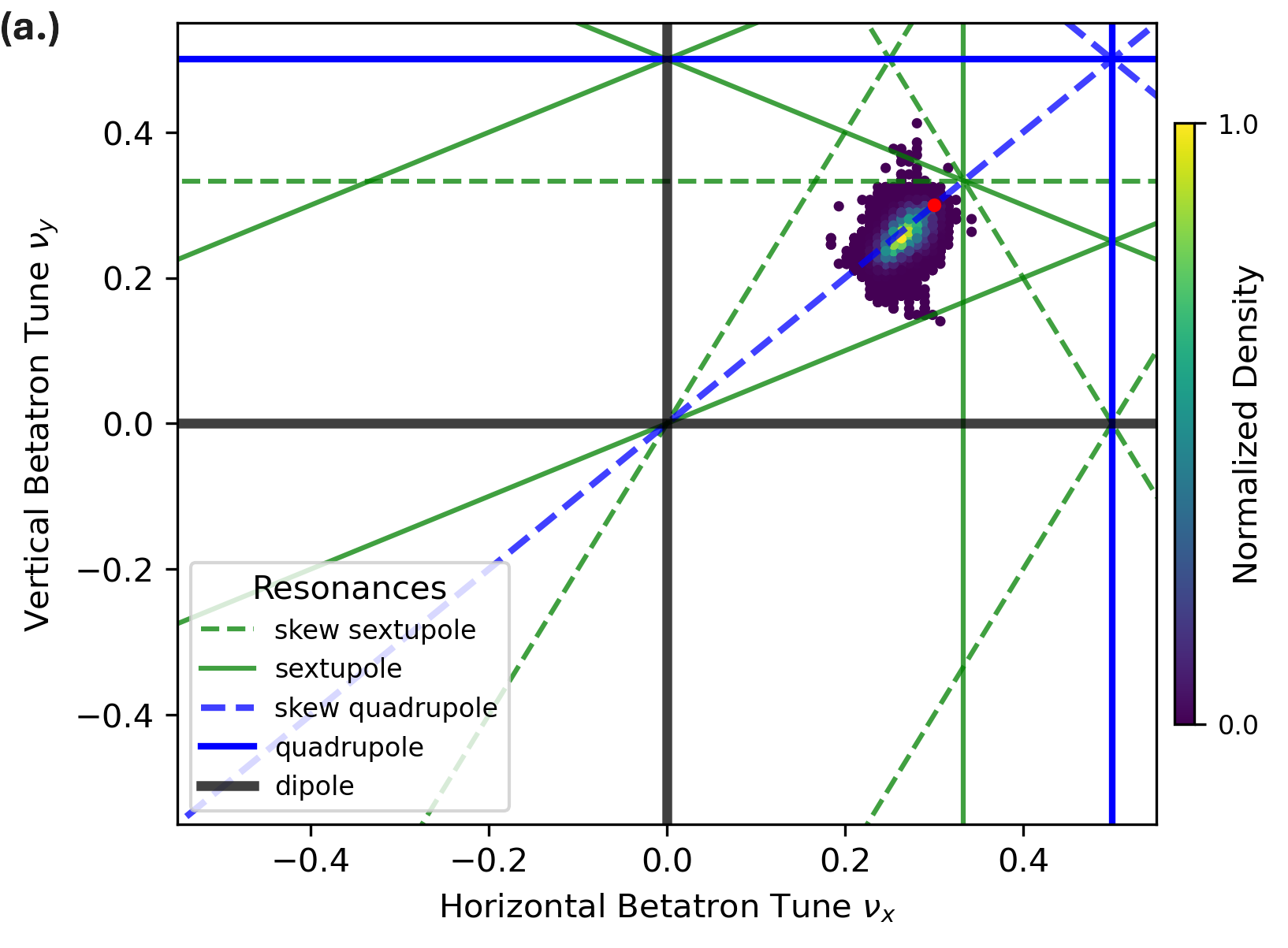} }}%
    \subfloat{{\includegraphics[width=0.49\columnwidth]{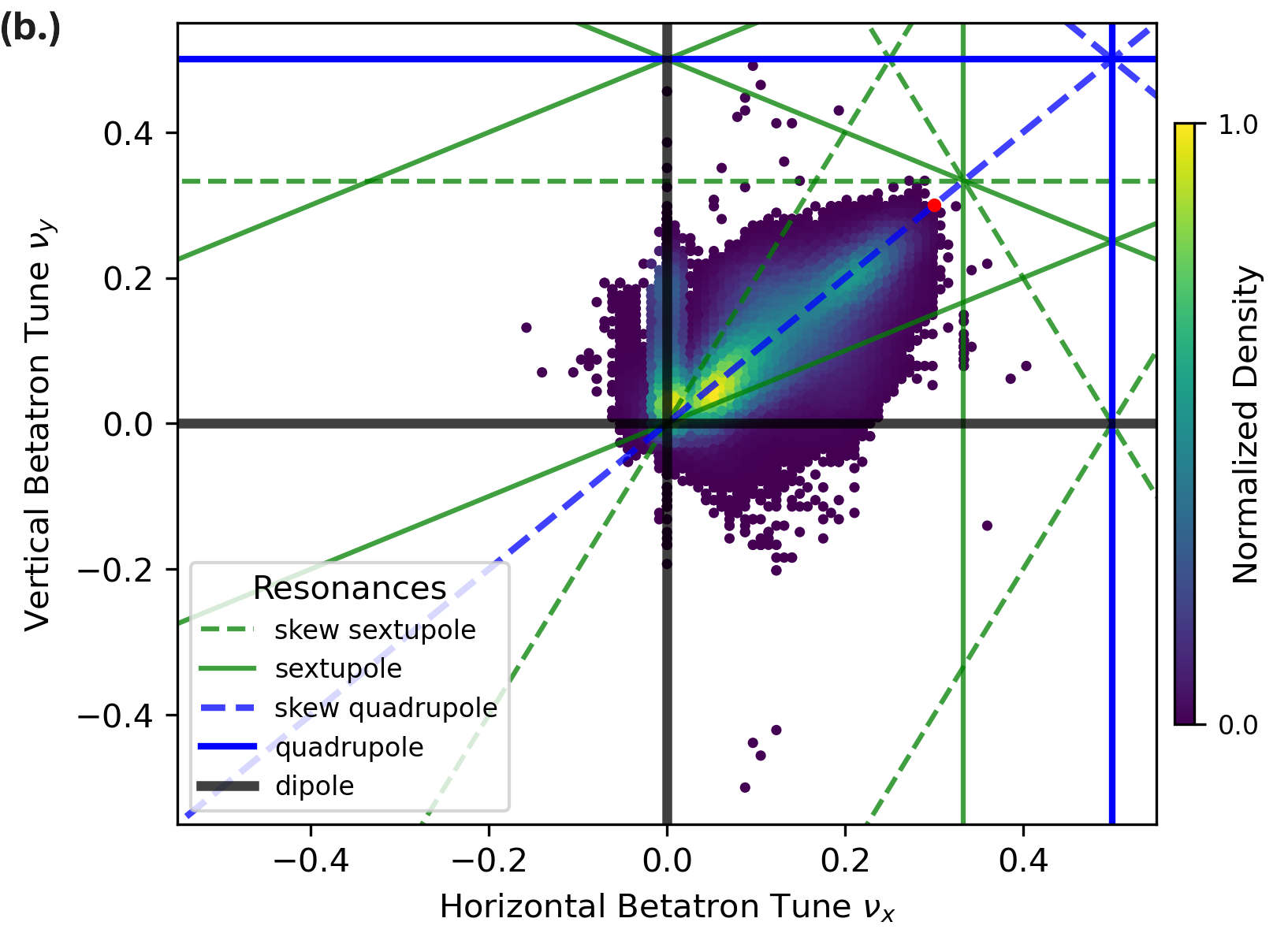} }}%
    \caption{Tune footprints for low current ($0.5$mA, panel a) and high current ($8$mA, panel b). Each data point represents the tune of a single macroparticle extracted via FFT from turn-by-turn positional data for 1 synchrotron period (114 turns) after adiabatic capture and before snap rotation. The red dot represents the fractional tune for the bare, linear lattice operating point of $Q_{x,y}=0.3$. The color indicates normalized density.}%
    \label{fig:Tunespreads}%
\end{figure}

The large space charge effect on the betatron tunes of the beam can be visualized by looking at the tune footprint of the beam, as seen in Fig.~\ref{fig:Tunespreads}.

In panel (a.) from Fig.~\ref{fig:Tunespreads}, we can already see that the effect of the space-charge force causes a smearing of the tune footprint, even at such a low intensity, pushing particles out of the core and closer to possibly dangerous resonances. Panel (b.) of the same figure shows the footprint of the 8~mA beam where we can now see a greater degree of smearing, causing the core of the beam to directly interact with the integer resonance, which inevitably leads to particle losses. It should be noted that the tune footprints shown in Fig.~\ref{fig:Tunespreads} show the measured tunes only for the time frame after capture but before the snap rotation. Looking back at Eq.~\ref{eq5}, the tune shift grows larger as the bunch length gets smaller, so the resulting tune shifts are nearly twice as extreme (if an accurate measurement could be made from the $\sim20$ turns of the snap rotation process).

\begin{figure}[!htb]%
    \centering
    \includegraphics[width=0.7\columnwidth]{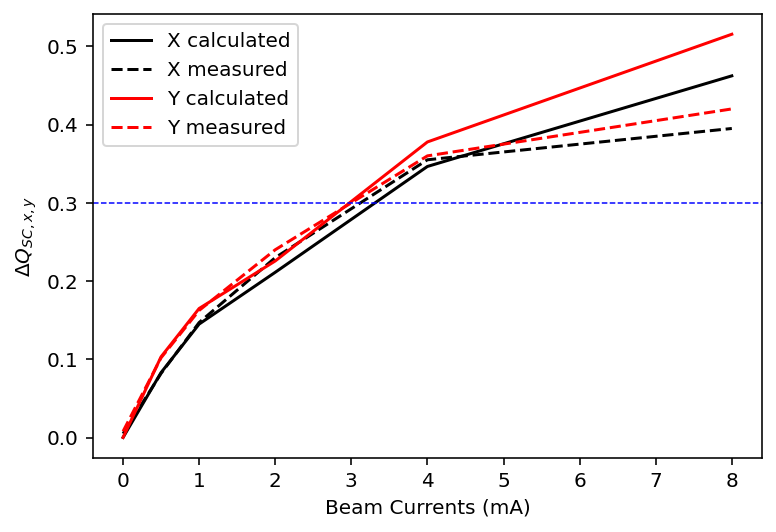}
    \caption{Comparison of space-charge tune shift parameters, calculated from Eq.~\ref{eq5} (solid) and measured from simulation (dashed) as functions of beam current. The blue horizontal line indicates a possible encounter with the integer resonance.}
    \label{fig:DQs}
\end{figure}

With ImpactX being a Particle-in-Cell (PIC) simulation code, tracking test particles with varying initial conditions in addition to the whole distribution can sometimes be unreliable, so instead, the tunes of all particles are measured and the particle with the smallest tune provides the value for measurement. A comparison of the measured tune shifts to the calculated tune shifts, only for the stationary RF, taken from Fig.~\ref{fig:DelQSC} is shown in Fig.~\ref{fig:DQs}.

\begin{figure}[!htb]%
    \centering
    \subfloat{{\includegraphics[width=0.49\columnwidth]{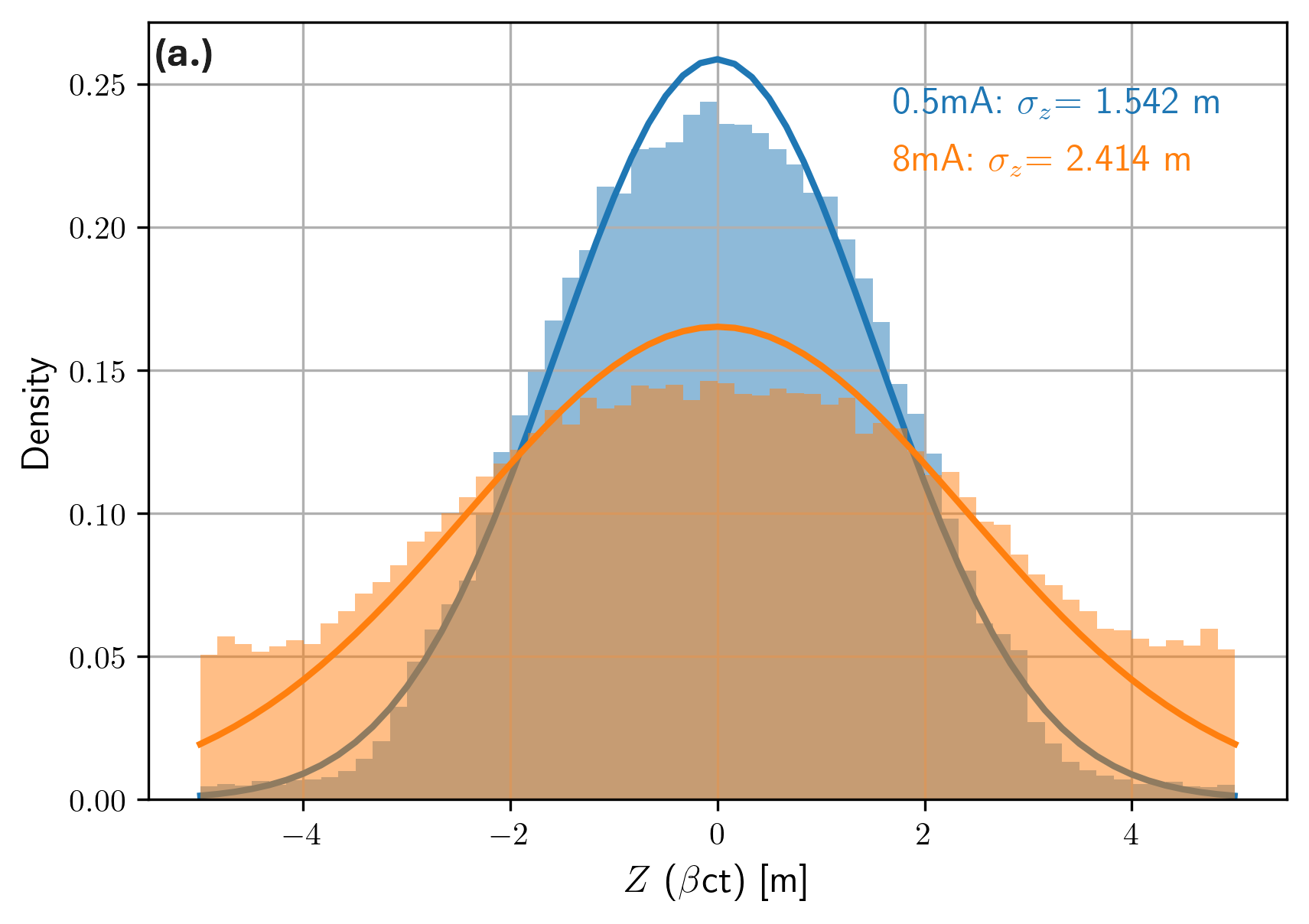} }}%
    \subfloat{{\includegraphics[width=0.49\columnwidth]{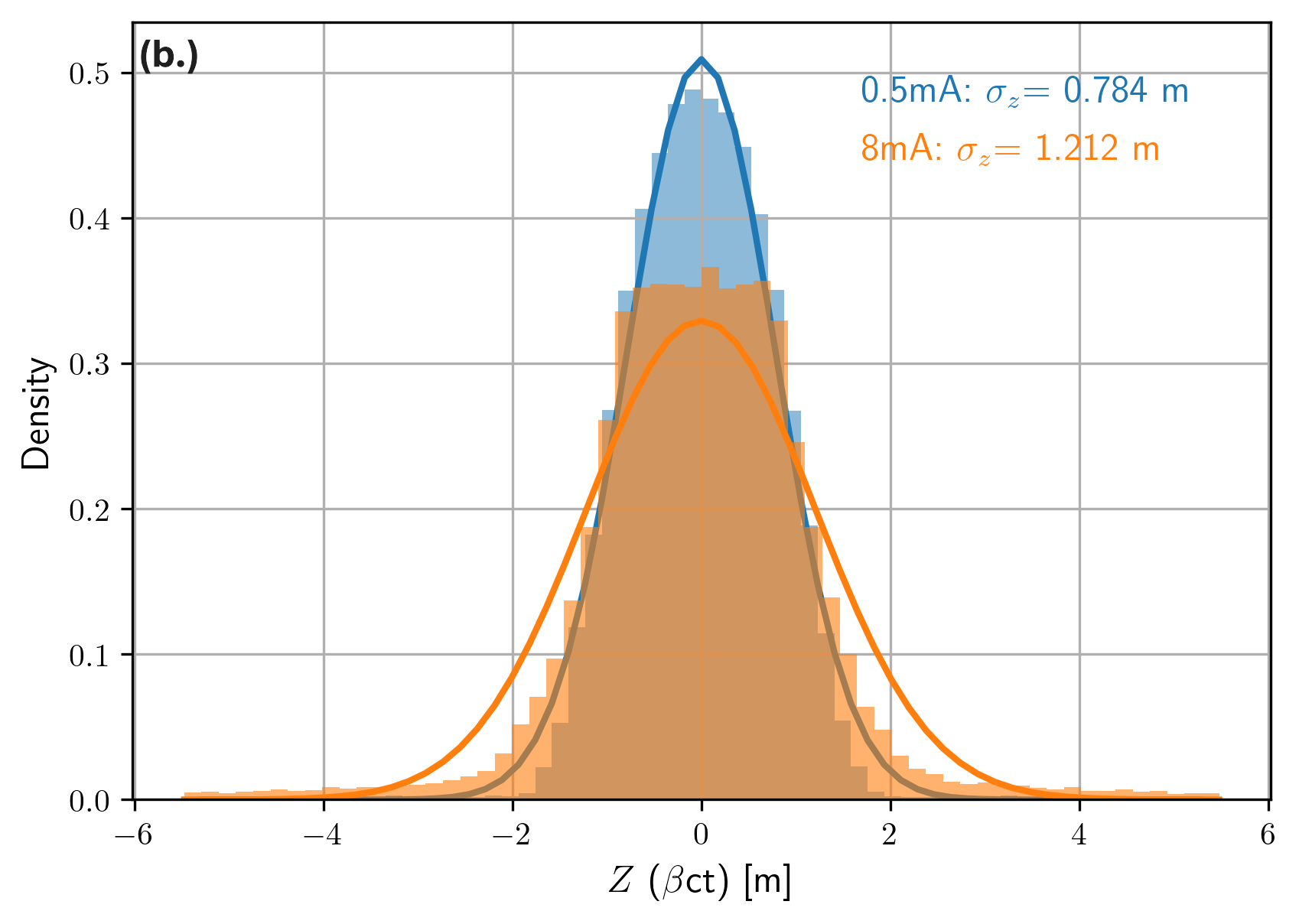} }}%
    \caption{Longitudinal density profiles for a low injected current (blue, $0.5$~mA) and a high injected current (orange, $8$~mA). Panel (a.) shows the distribution after adiabatic capture and panel (b.) shows the distribution at its most compressed bunch length after the snap rotation.}%
    \label{fig:DensityProfiles}%
\end{figure}

\begin{figure}[!htb]
    \centering
    \includegraphics[width=1.0\columnwidth]{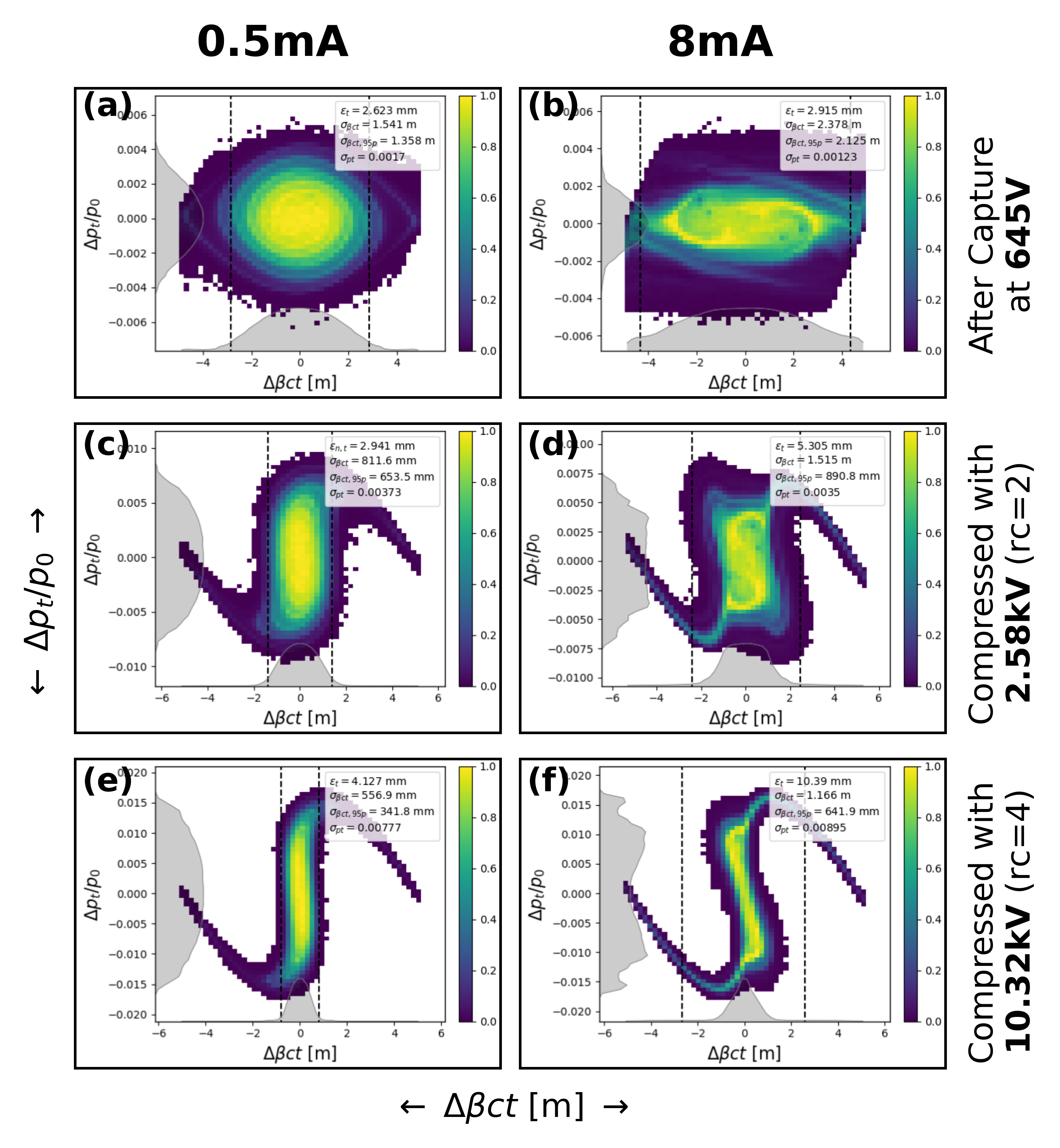}
    \caption{Longitudinal phase spaces for low current ($0.5$mA) in the left column and high current ($8$mA) in the right column. Row one shows the longitudinal phase spaces just before snap rotation and rows two and three show the longitudinal phase spaces when the bunch is most compressed for two different compression factors, where $r_c$ is the compression factor defined in Eq.~\ref{eq4}.}
    \label{fig:ADCap_Rot}
\end{figure}

At the highest injected current of 8~mA, a noticeable divergence emerges between the calculated and measured tune shifts. While the calculated values from Eq.~\ref{eq5} continue to scale predictably with current, the measured tune shifts - extracted from the particle-resolved FFT analysis mentioned previously - fall short of the expected depression. This discrepancy likely stems from the nonlinear deformation of the RF bucket seen by the bunch under extreme space-charge forces. As we will see in Figs.~\ref{fig:DensityProfiles} and~\ref{fig:ADCap_Rot}, the longitudinal profile at 8~mA becomes significantly distorted, with the core density flattened and the bunch length broadened. This redistribution dilutes the peak space-charge density and alters the transverse focusing environment, especially for particles near the tails. Additionally, the measured tune shifts reflect the minimum tune among all of the macroparticles, which may be dominated by particles that have migrated outward due to defocusing, rather than those remaining in the high-density core. The result is a measured tune shift that under-represents the true core depression predicted by the analytical model. This behavior highlights the limitations of using single-particle FFT metrics in regimes where the beam undergoes strong space-charge-driven redistribution and suggests that more robust diagnostics may be necessary for accurate characterization at high intensity, a topic that will be explored in future works for this project. Barring that caveat, excluding 8~mA, the calculated tune shifts and minimum particle-resolved tunes from the FFT analysis, as seen in Fig.~\ref{fig:DQs}, show excellent agreement. This validates the space-charge tune depressions reported in Fig.~\ref{fig:DelQSC}, and confirms the parameters of the space-charge simulation are well understood.

Figure~\ref{fig:DensityProfiles} shows the longitudinal density profiles, and their corresponding Gaussian fit curves, for a case of low beam current and a case of high beam current.
The low current distribution shown in Fig.~\ref{fig:DensityProfiles} retains a clear Gaussian shape with a distinct peak, both after capture and snap rotation. In contrast, the high current case shows significant core flattening, especially in panel (b.), where the compressed bunch more closely resembles a nearly uniform distribution. This profile distortion reflects the strong longitudinal space charge forces at high intensity, which broaden the core and completely flatten the peak charge.

Figure~\ref{fig:ADCap_Rot} shows the longitudinal phase-space distributions for the case of a low injected current (left column) and a high injected current (right column). The first row, panels (a)-(b), show the longitudinal phase space after capture and one synchrotron period (1396 turns), for a capture voltage of $645$~V. The second row, panels (c)-(d), show the beam at its most compressed after snap rotating at $2.58$~kV ($r_c=2$). The last row, panels (e)-(f), show the beam at its most compressed after snap rotating at $10.32$~kV ($r_c=4$). 
The two vertical dotted lines in each panel are outlining the region in which 95\% of the beam is located. Shown in Fig.~\ref{fig:ADCap_Rot}, for $8$~mA, there is a significant distortion to the RF bucket shape as a result of the longitudinal space charge forces diluting the core density.

\begin{figure}[!htb]
    \centering
    \includegraphics[width=0.75\columnwidth]{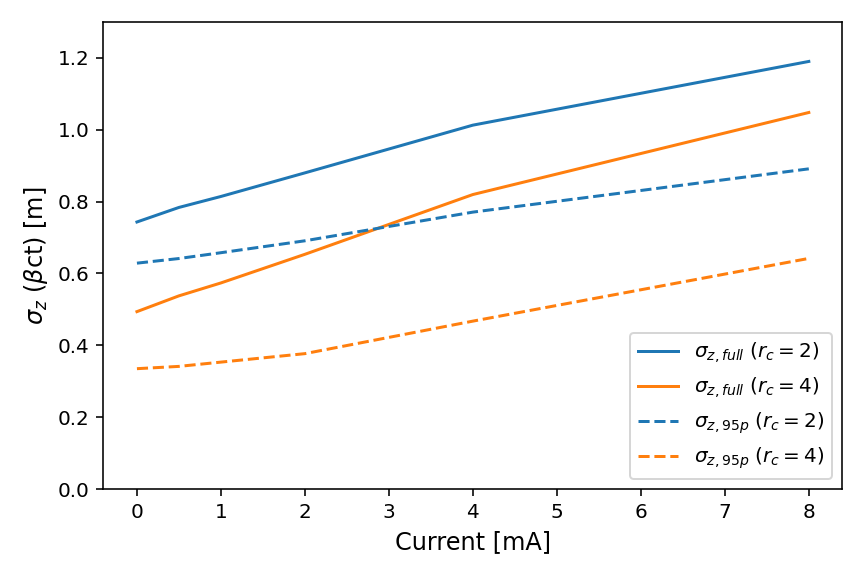}
    \caption{Final, compressed, rms bunch lengths as a function of injected current for $r_c=2$ (blue) and $r_c=4$ (orange). The solid lines are the rms bunch lengths of the full bunch and the dotted lines are the rms bunch lengths of the central 95\% of the bunch.}
    \label{fig:RCs}
\end{figure}

Figure~\ref{fig:RCs} shows the evolution of the longitudinal rms beam size as a function of the beam current.  The solid lines represent the rms values of the entire bunch and the dotted lines represent $1\sigma$ of the central 95\% of the bunch, capturing mainly the core of the distribution. The $r_c=2$ (blue) values correspond to capturing at $645$~V and then snap rotating at $2.58$~kV and the $r_c=4$ (orange) values correspond to capturing at $645$~V and snap rotating at $10.32$~kV.

A clear positive correlation is observed, indicating that longitudinal broadening intensifies with current, consistent with space charge induced lengthening. The divergence between the total and central 95\% rms curves suggests that the beam tails contribute increasingly to the overall size at higher currents, highlighting the importance of non-Gaussian features in longitudinal distribution modeling. From Fig.~\ref{fig:RCs}, we conclude that the final achievable bunch length depends on the injected beam intensity. 

\begin{figure}[!htb]
    \centering
    \includegraphics[width=0.75\columnwidth]{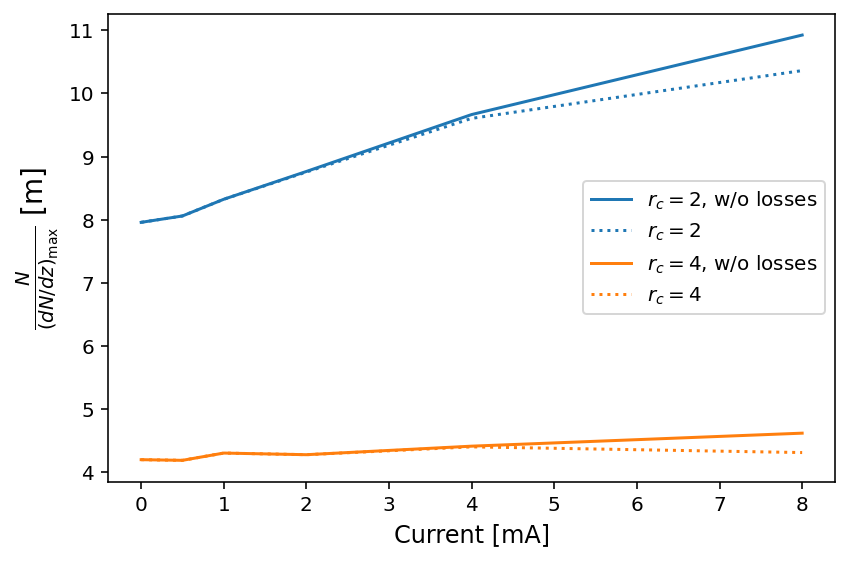}
    \caption{The ratio of total particles over the peak line charge density of the bunch as a function of injected current. The solid lines represent idealized cases, without any losses, while the dotted lines show the impact of including losses due to the beam pipe apertures.}
    \label{fig:Prots_over_maxLineCharge}
\end{figure}

\begin{figure}[!htb]
    \centering
    \includegraphics[width=0.75\columnwidth]{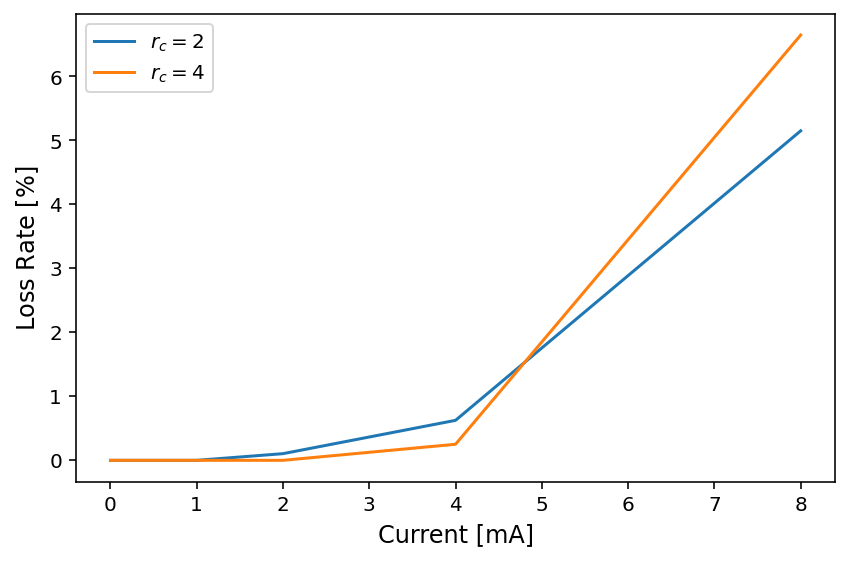}
    \caption{Loss rate as a function of beam current for the two different compression scenarios. A reminder that the only apertures included represent the 2-inch diameter vacuum beam pipe. For all simulations, most of the losses occur at the end, during snap rotation, except for 8~mA, where most of the losses occur during the RF capture.}
    \label{fig:LossComp}
\end{figure}

To complement the rms bunch length analysis from Fig.~\ref{fig:RCs}, Fig.~\ref{fig:Prots_over_maxLineCharge} presents $N/(dN/dz)_\text{max}$ as a function of injected beam current for the two compression scenarios. 
As can be seen from Fig.~\ref{fig:Prots_over_maxLineCharge}, the larger compression ratio is more effective in creating larger longitudinal charge density at higher tune shifts, while generating only slightly more losses, which is emphasized in Fig.~\ref{fig:LossComp}. Figure~\ref{fig:Prots_over_maxLineCharge} essentially provides a measure of compactness, with a smaller value indicating that the bunch is more sharply peaked, as is the case with the factor of four compression scheme. 

Figure~\ref{fig:LossComp} highlights the interplay between compression and loss dynamics. At low currents, both compression scenarios yield negligible losses. As current increases, space-charge forces intensify, amplifying halo-driven losses. The comparison underscores the balance act between aggressive compression and an increased sensitivity to intensity-dependent degradation with implications for both beam quality and operational thresholds. 

\section{Conclusion and Outlook} 

While the generation of short, high-intensity proton bunches is critical for many accelerator-based particle physics facilities, it faces several challenges and benefits from both theoretical insight and experimental validation. For the proton driver of a future 10+ TeV Muon Collider \cite{JeffIPAC,SophiaIPAC}, efficiently shortening the proton pulse in a Compressor Ring will be a critical part of achieving meaningful collider luminosity. The performance of that bunch compression process will be driven by managing the extreme longitudinal and transverse space-charge forces.

We have proposed the FAST/IOTA Bunch Rotation Experiment (FIBRE), to demonstrate the compression of intense proton beams into short pulses at FAST/IOTA and enhance the technology for next generation proton drivers. FIBRE provides a platform to validate our beam dynamics simulations and to conduct detailed studies of the interplay between a lattice's phase-slip factor and the compression process in a space-charge-dominated proton beam, including transverse and longitudinal beam dynamics, emittance growth, and potential countermeasures to mitigate it. Support for the FIBRE experiment at IOTA only requires an increase in the available RF voltage at 2.2~MHz, and appropriate RF technology can achieve bunch compression by a factor of two or more.


In our computer simulations using the ImpactX code, we applied various RF manipulation schemes to optimize bunch compression and studied in detail the evolution of the 6D charge distribution and rms emittances across the operational current range (0–8 mA). Our results demonstrate the significant influence of both transverse and longitudinal space-charge forces, characterized by a tune-shift parameter of $-\Delta Q_{SC} \ge 0.5$ and a longitudinal space-charge defocusing that can become comparable in magnitude to the focusing effect produced by the RF voltage. The parameters of the FAST/IOTA ring highlight the advantage of operating with a relatively large phase-slip factor $\eta$, suggesting a promising design direction for future muon collider compressor rings.

Next steps include continued simulations with ImpactX, focusing on optimizing the RF system under realistic hardware constraints and exploring bunch compression factors exceeding two. To mitigate longitudinal space-charge effects during compression, we plan to investigate the use of inductive inserts, as proposed in~\cite{IndIns} and experimentally demonstrated at the LANL Proton Storage Ring~\cite{IndInsPSR}.

\acknowledgments
We are thankful to Chad Mitchell of LBNL for help with installation of ImpactX and consulting on the use of the code. This work used resources of the Center for Research Computing and Data at Northern Illinois University.

This work was produced by FermiForward Discovery Group, LLC under Contract\\
No. 89243024CSC000002 with the U.S. Department of Energy, Office of Science, Office of High Energy Physics. Publisher acknowledges the U.S. Government license to provide public access under the DOE Public Access Plan.




\bibliographystyle{JHEP}
\bibliography{mbiblio}

\end{document}